\documentclass[a4paper,11pt]{article}
\pdfoutput=1 

\usepackage{jcappub} 

\usepackage[T1]{fontenc} 

\usepackage{aas_macros}
\usepackage{xspace}
\usepackage{bm}

\newcommand{\eg}{\mbox{e.\,g.,}\xspace}

\newcommand{\ie}{\mbox{i.\,e.,}\xspace}
\newcommand{\Ie}{\mbox{I.\,e.,}\xspace}
\renewcommand{\vec}[1]{\bm{#1}}

\definecolor{red}{rgb}{1,0,0}

\title{Bayesian emulator optimisation for cosmology: application to the Lyman-alpha forest}

\author[a,1]{Keir K. Rogers,\note{Corresponding author.}}
\author[b, a]{Hiranya V. Peiris,}
\author[b]{Andrew Pontzen,}
\author[c]{Simeon Bird,}
\author[d, e]{Licia Verde}
\author[b]{and Andreu Font-Ribera}

\affiliation[a]{Oskar Klein Centre for Cosmoparticle Physics, Stockholm University,\\AlbaNova, Stockholm SE-106 91, Sweden}
\affiliation[b]{Department of Physics \& Astronomy, University College London,\\Gower Street, London WC1E 6BT, UK}
\affiliation[c]{Department of Physics \& Astronomy, University of California, Riverside,\\900 University Avenue, Riverside, CA 92521, USA}
\affiliation[d]{Institut de Ci\`encies del Cosmos, University of Barcelona,\\ICCUB, Barcelona 08028, Spain}
\affiliation[e]{Instituci\'o Catalana de Recerca i Estudis Avan\c{c}ats,\\Passeig Llu\'is Companys 23, Barcelona 08010, Spain}

\emailAdd{keir.rogers@fysik.su.se}
\emailAdd{h.peiris@ucl.ac.uk}
\emailAdd{a.pontzen@ucl.ac.uk}
\emailAdd{sbird@ucr.edu}
\emailAdd{liciaverde@icc.ub.edu}
\emailAdd{a.font@ucl.ac.uk}

\abstract{The Lyman-alpha forest provides strong constraints on both cosmological parameters and intergalactic medium astrophysics, which are forecast to improve further with the next generation of surveys including eBOSS and DESI. As is generic in cosmological inference, extracting this information requires a likelihood to be computed throughout a high-dimensional parameter space. Evaluating the likelihood requires a robust and accurate mapping between the parameters and observables, in this case the 1D flux power spectrum.  Cosmological simulations enable such a mapping, but due to computational time constraints can only be evaluated at a handful of sample points; ``emulators'' are designed to interpolate between these. The problem then reduces to placing the sample points such that an accurate mapping is obtained while minimising the number of expensive simulations required.  To address this, we introduce an emulation procedure that employs Bayesian optimisation of the training set for a Gaussian process interpolation scheme. Starting with a Latin hypercube sampling (other schemes with good space-filling properties can be used), we iteratively augment the training set with extra simulations at new parameter positions which balance the need to reduce interpolation error while focussing on regions of high likelihood. We show that smaller emulator error from the Bayesian optimisation propagates to smaller widths on the posterior distribution. Even with fewer simulations than a Latin hypercube, Bayesian optimisation shrinks the 95\% credible volume by 90\% and, \eg the \(1 \sigma\) error on the amplitude of small-scale primordial fluctuations by 38\%. This is the first demonstration of Bayesian optimisation applied to large-scale structure emulation, and we anticipate the technique will generalise to many other probes such as galaxy clustering, weak lensing and 21cm.}

\begin{document}
\maketitle
\flushbottom

\section{Introduction}
\label{sec:intro}

The cosmic large-scale structure informs us about the late-time evolution of the Universe as well as bearing imprints of the primordial fluctuations; the most accurate modelling of this epoch requires numerical simulation. For example, the Lyman-alpha forest is simultaneously sensitive to a wide range of cosmological and astrophysical parameters. It is sourced by the (mildly) non-linear gas in the intergalactic medium (IGM), meaning that forward modelling requires the computationally-expensive calculation of a cosmological hydrodynamical simulation. This requires one to estimate the likelihood function with \emph{millions} of samples (to adequately sample the parameter space by Markov chain Monte Carlo methods) while only being able to compute \emph{tens} of cosmological simulations.

This challenge is worth overcoming because of the unique range of scales (hundreds of Mpc to sub-Mpc) and redshifts (\(2 < z < 5\)) at which the Lyman-alpha forest is observed. This allows the forest to put tight limits on the presence of additional cosmological components like massive neutrinos or non-cold dark matter, and deviations from a power-law primordial power spectrum (on small scales) \cite{2005PhRvD..71j3515S, 2015JCAP...11..011P, 2017arXiv170201764I, 2017arXiv170203314Y, 2017arXiv170304683I, 2017arXiv170309126A}; and (on large scales) measure the cosmological expansion rate \cite{2011JCAP...09..001S, 2013A&A...552A..96B, 2013JCAP...03..024K, 2013JCAP...04..026S, 2015A&A...574A..59D, 2017arXiv170200176B} and geometry \cite{1979Natur.281..358A, 1999ApJ...511L...5H, 1999ApJ...518...24M, 2003ApJ...585...34M} at a redshift before dark energy came to dominate the energy contents of the Universe. The Lyman-alpha forest is also sensitive to the thermal history (temperature and density) of the IGM from the end of hydrogen reionisation through the reionisation of helium, \eg refs.~\cite{1998MNRAS.298L..21H, 2018arXiv180804367W}. These constraints are forecast \cite{2014JCAP...05..023F} to tighten further with ongoing and future spectroscopic surveys, \eg the extended Baryon Oscillation Sky Survey (eBOSS) \cite{2016AJ....151...44D} and the Dark Energy Spectroscopic Instrument (DESI) \cite{2016arXiv161100036D, 2016arXiv161100037D}. It may also be possible to learn about the sources of ionising radiation on large scales \cite{2014ApJ...792L..34P}. However, in order to learn about either cosmology or astrophysics, it is necessary to marginalise over the uncertainty in the other. It follows that in order to estimate the likelihood function of a given dataset, it is necessary to simultaneously vary multiple parameters (\eg in this study, we consider a model with seven parameters in total). This puts even further strain on the small number (approximately 50 -- 100; \eg ref.~\cite{2018arXiv180804367W} run a suite of \(\sim 70\)) of cosmological simulations that can be reasonably run in the available computing time.

The solution lies in \emph{emulating} the outputs of simulations. This is a form of interpolation, meaning that a small number of forward simulations can be used to predict simulation outputs throughout parameter space. Emulators have found use in various branches of science, wherever forward modelling is computationally expensive due \eg either to complex non-linearities (\eg fluid dynamics problems in engineering \cite{Sacks_1989, Queipo_2005, forrester2008engineering, forrester2009recent}) or an extremely large number of elements that require calculation (\eg the simulation of microbial communities in biology \cite{OYEBAMIJI201769}).

Indeed, emulators have found use in modelling the cosmological large-scale structure, \eg modelling the small-scale non-linear matter power spectrum \cite{Heitmann:2009}; the galaxy power spectrum \cite{Kwan:2013, 2018arXiv180405867Z}; galaxy weak lensing peak counts and power spectrum \cite{Liu:2015, Petri:2015}; the 21cm power spectrum \cite{2018arXiv181109141J}; or the halo mass function \cite{2018arXiv180405866M}. Ref.~\cite{2018arXiv180804367W} emulate the one-dimensional (1D) Lyman-alpha forest flux power spectrum for three thermal parameters and the mean flux in their inference on the IGM thermal history (for a fixed cosmology). Also, ref.~\cite{2018PhRvD..98h3540M} interpolate the small-scale (\(0.005\,\mathrm{s}\,\mathrm{km}^{-1} \leq k_{||} \leq 0.08\,\mathrm{s}\,\mathrm{km}^{-1}\)), high-redshift (\(4.2 \leq z \leq 5\)) 1D flux power spectrum using the ``ordinary kriging'' method (\eg ref.~\cite{webster2007geostatistics}) for the study of non-cold dark matter models.

The standard approach for the interpolation of Lyman-alpha forest observables (as used in refs.~\cite{2006MNRAS.365..231V, 2011MNRAS.413.1717B, 2013A&A...559A..85P, 2015JCAP...02..045P}) is to make use of quadratic polynomial interpolation using a second-order Taylor expansion around a fiducial simulation. However, most importantly for statistical inference, this method gives no theoretical error estimate in the interpolation. Instead, errors are estimated empirically from test simulations and a global worst-case error is assigned. In this work and a companion article \cite{emulator_forest_GP_2018}, we use Gaussian processes to model the 1D flux power spectrum for a full set of cosmological and astrophysical parameters. A Gaussian process is a stochastic process where any finite subset forms a multivariate Gaussian distribution (see ref.~\cite{gpml} for a review). The Gaussian process provides a principled theoretical estimate of the uncertainty in interpolated simulation values, eliminating any need to resort to worst-case empirical estimates.

The key practical element is the construction of the training dataset for optimising the Gaussian process. In this study, we investigate Bayesian optimisation \cite{kushner1964new, mockus1978toward, Mockus1994, 10.2307/2673557, kennedy2001bayesian} as a method to decide where in parameter space to run training simulations. In cosmology, ref.~\cite{2018PhRvD..98f3511L} used Bayesian optimisation in likelihood-free inference from Joint Light-curve Analysis supernovae data \cite{2014A&A...568A..22B} in order to gain more precise posterior distributions with fewer forward simulations than existing methods. The approach iteratively uses knowledge about the approximate likelihood function and the regions of parameter space where there is greatest uncertainty. The principle of Bayesian optimisation is to propose new training samples balancing \emph{exploration} of the prior volume where the current uncertainty in the optimised function is highest with \emph{exploitation} of previous iterations of the emulator revealing the most interesting (for us, high-likelihood) regions. Bayesian optimisation has been developed as a technique to find the optima of functions in as few evaluations as possible; it achieves this by using prior information in the manner set out above. This naturally dovetails into Gaussian process emulation, which provides a robust estimate of uncertainty in predictions across the parameter space \cite{Locatelli1997}. The details of the balance between exploration and exploitation are encoded in the acquisition function used to determine future proposals, of which many examples have been developed (we use a novel expansion of the GP-UCB acquisition function) \cite{Cox97sdo:a, auer2002using, auer2002finite, dani2008stochastic, 2017arXiv170400520J}. We also show how to propose multiple training samples simultaneously (batch acquisition).

The key point here is that cosmologists are mostly interested in accurately characterising the peak of the posterior distribution (\ie the \(\sim 95 \%\) -- \(99 \%\) credible region). Indeed, the standard approach to sampling the posterior distribution -- Markov chain Monte Carlo (MCMC) methods -- is optimised for this purpose. Therefore, we do not actually require (and often cannot afford) uniform accuracy in the modelling of cosmological observables across the prior volume. Bayesian optimisation solves this problem by concentrating available resources in regions of high posterior probability using informed Bayesian decision making.

Bayesian optimisation can help in our inference problem, where we specify the likelihood function but where model evaluations are extremely computationally expensive and the parameter space is high-dimensional -- as well as many other inference problems in cosmology with computationally expensive forward simulations and many parameters. This is compared to the ``brute-force'' method of simply increasing the number of simulations in the uniform Latin hypercube sampling scheme described in our companion paper \cite{emulator_forest_GP_2018}. This could ``waste'' samples on the edges of the prior volume. The informed decision-making of the Bayesian optimisation can ultimately lead to the robust statistical inference necessary to exploit the full potential of ongoing and future spectroscopic surveys such as eBOSS and DESI -- as well as improve the performance of many other cosmological emulators.

The article is organised as follows. In section~\ref{sec:gaussian_process}, we review the use of Gaussian processes to emulate the 1D Lyman-alpha forest flux power spectrum. We detail our use of Bayesian optimisation in section~\ref{sec:optimisation}. We present our main results in section~\ref{sec:results}, discuss them in section~\ref{sec:discussion} and draw our final conclusions in section~\ref{sec:conclusions}.

\section{Method}
\label{sec:method}

\subsection{Gaussian process emulator}
\label{sec:gaussian_process}

The full details of our Gaussian process emulator are given in our companion paper \cite{emulator_forest_GP_2018}; here we summarise the key points.

\subsubsection{Simulated training data}
\label{sec:training}
The data vector which we emulate (see section~\ref{sec:interpolation}) and from which we calculate our likelihood function (see section~\ref{sec:likelihood}) is the 1D Lyman-alpha forest flux power spectrum \(P^\mathrm{1D}(\vec{\theta}; k_{||}, z)\) -- line-of-sight fluctuations of transmitted flux in quasar spectra. It is a function of line-of-sight wavenumber \(k_{||}\), redshift \(z\) and cosmological and astrophysical model parameters \(\vec{\theta}\). In order to evaluate accurate theoretical predictions for \(P^\mathrm{1D}\), it is necessary to run a cosmological hydrodynamical simulation for each  \(\vec{\theta}\). For this, we use the publicly-available code \texttt{MP-Gadget}\footnote{\url{https://github.com/MP-Gadget/MP-Gadget}} \cite{yu_feng_2018_1451799}, itself derived from the public \texttt{GADGET-2} code \cite{2001NewA....6...79S, 2005MNRAS.364.1105S}, in order to evolve \(256^3\) particles each of dark matter and gas in a \((40\,h^{-1}\,\mathrm{Mpc})^3\) box\footnote{Our hydrodynamical simulations lack the size or resolution for a robust comparison to real data, but they are sufficient for the explication of this method. We do not expect any complication in applying this method with fully-converged simulations; the only difference is that simulated flux power spectra will be more precisely determined. In particular, the Gaussian process model and Bayesian optimisation method that we prove here will be equally applicable to these converged simulations.} from a starting redshift \(z = 99\) to the redshifts at which the Lyman-alpha forest is observed in BOSS Data Release 9 (DR9) \cite{2011AJ....142...72E, 2013AJ....145...10D, 2013A&A...559A..85P}, \(2.2 \leq z \leq 4.2\). From the simulated particle data, we then generate mock quasar spectra containing only the Lyman-alpha absorption line and calculate the 1D flux power spectrum using \texttt{fake\_spectra} \cite{2017ascl.soft10012B}. We use the same 35 \(k_{||}\) and 11 \(z\) bins as used in BOSS DR9.

The model parameters \(\vec{\theta}\) we elect to use are:
\begin{itemize}
\item three cosmological: the amplitude \(A_\mathrm{s}\) and scalar spectral index \(n_\mathrm{s}\) of the primordial power spectrum with a pivot scale \(k_\mathrm{pivot} = \frac{2 \pi}{8}\,\mathrm{Mpc}^{-1}\) corresponding to the scales probed by the 1D flux power spectrum, and the dimensionless Hubble parameter\footnote{Note that we otherwise fix the (physical) total matter density \(\Omega_\mathrm{m} h^2 = 0.1327\) and baryon density \(\Omega_\mathrm{b} = 0.0483\). Indeed, the vast majority of our apparent constraining power on \(h\) in section~\ref{sec:results} arises from the inverse scaling of \(\Omega_\mathrm{m}\) and its effect on the linear growth factor.} \(h\);
\item two astrophysical: the heating amplitude \(H_\mathrm{A}\) and slope \(H_\mathrm{S}\) of the (over)density \(\Delta\) - dependent rescaling of photo-heating rates\footnote{This effectively allows us to modify the temperature-density relationship of the IGM gas \cite{2008MNRAS.386.1131B}.} \(\epsilon = H_\mathrm{A} \epsilon_0 \Delta^{H_\mathrm{S}}\);
\item two for the mean flux, a multiplicative correction to the amplitude \(\tau_0\) and an additive correction to the slope \(\mathrm{d}\tau_0\) of the empirical redshift dependence of the mean optical depth as measured by ref.~\cite{Kim:2007}.
\end{itemize}
We use a uniform prior (in the evaluation of our posterior distribution of model parameters; see section~\ref{sec:likelihood}) on \(\vec{\theta}\) as given in table~\ref{tab:prior} (the limits differ slightly from our companion paper \cite{emulator_forest_GP_2018}). More details about our parameterisation and possible alternatives are discussed in our companion paper \cite{emulator_forest_GP_2018}.

\begin{table}[tbp]
\centering
\begin{tabular}{|l|c|}
\hline
Model parameter&Prior range\\
\hline
\(A_\mathrm{s}\) & \([1.5 \times 10^{-9}, 2.8 \times 10^{-9}]\)\\
\(n_\mathrm{s}\) & \([0.9, 0.99]\)\\
\(h\) & \([0.65, 0.75]\)\\
\(H_\mathrm{A}\) & \([0.6, 1.4]\)\\
\(H_\mathrm{S}\) & \([-0.4, 0.4]\)\\
\(\tau_0\) & \([0.75, 1.25]\)\\
\(\mathrm{d}\tau_0\) & \([-0.25, 0.25]\)\\
\hline
\end{tabular}
\caption{\label{tab:prior}The ranges of the uniform prior on our model parameters.}
\end{table}

Before considering Bayesian optimisation (see section~\ref{sec:optimisation}), the basic set-up of our emulator is to generate training data (the sample points from which we emulate) on a Latin hypercube sampling of the full prior volume. We calculate \(P^\mathrm{1D}(\vec{\theta}; k_{||}, z)\) as described above at a certain number \(N_\mathrm{Latin}\) (our base emulator uses 21 training samples) of values of \(\vec{\theta}\), as sampled by the Latin hypercube. A Latin hypercube is a random sampling scheme with good space-filling properties (low discrepancy). It sub-divides the prior volume along each axis into \(N_\mathrm{Latin}\) equal sub-spaces. It then randomly distributes \(N_\mathrm{Latin}\) samples under the constraint that each sub-space is sampled once. We randomly generate many different Latin hypercubes and choose the one that maximises the minimum Euclidean distance between different samples.

The number of simulations necessary in the base Latin hypercube emulator for accurate estimation of the likelihood function is a non-trivial choice dependent on the number of model parameters, the size of the prior volume and the particular correlation structure of the parameter space -- \ie quite particular to the problem at hand. In our testing, however, we found that if the density of training samples in the full prior volume \(N_\mathrm{Latin} / V_\mathrm{prior}\) is insufficient, the resulting large error in the interpolation (same order of magnitude as the ``measurement'' error) causes spurious bias and/or multi-modality in the approximate likelihood function (see sections~\ref{sec:interpolation} and \ref{sec:likelihood}). Our base emulator of 21 simulations gives a good first approximation of the likelihood for this particular context (with the emulator-to-measurement error ratio \(\sim 20 \%\)); see section~\ref{sec:discussion} and our companion paper \cite{emulator_forest_GP_2018} for further discussion.

Because the mean flux can be adjusted in the post-processing of hydrodynamical simulations (with comparatively negligible computational cost), its parameters are handled differently. Although we sample in the likelihood function the two parameters \([\tau_0, \mathrm{d}\tau_0]\) described above, we actually emulate the mean optical depth separately in each redshift bin. This means that the \(N_\mathrm{Latin}\) training samples from above are only distributed in the five cosmological and astrophysical parameters. Then, at each of these training points, the mean optical depth in each redshift bin is sampled uniformly from a prior range translated from the prior ranges on \([\tau_0, \mathrm{d}\tau_0]\). We use 10 mean flux samples per redshift bin; this number can in effect be arbitrarily increased though tests show no improvement in the accuracy of Gaussian process modelling from doing so\footnote{Our treatment of the mean flux is somewhat akin to the concept of ``fast'' and ``slow'' parameters in MCMC sampling \cite{2002PhRvD..66j3511L}.}.

Finally, we anticipate sections~\ref{sec:likelihood} and \ref{sec:results} and note that we test our method on simulated rather than real data because we wish to compare our results to a known truth. We therefore generate a mock data vector \(\vec{d} \equiv P^\mathrm{1D}(\vec{\theta_\mathrm{true}}; k_{||}, z)\) by the same process as our theory evaluations described above, where \([\tau_0, \mathrm{d}\tau_0, A_\mathrm{s}, n_\mathrm{s}, h, H_\mathrm{A}, H_\mathrm{S}] = [0.95, 0, 2.24 \times 10^{-9}, 0.974, 0.685, 1.09, 0.0509]\).

\subsubsection{Gaussian process emulation as interpolation}
\label{sec:interpolation}
In section~\ref{sec:likelihood}, we will want to estimate our likelihood function by evaluating \(P^\mathrm{1D}(\vec{\theta}; k_{||}, z)\) for very many (\(\sim 10^6\)) different values of \(\vec{\theta}\), but we have only \(N_\mathrm{Latin} = 21\) model evaluations. We want to find a flexible model to \emph{interpolate} \(P^\mathrm{1D}(\vec{\theta}; k_{||}, z)\) between the evaluations we have and to have a robust estimate of the uncertainty in this interpolation so that we can include it in the statistical model. We achieve this by modelling the simulation output as a Gaussian process (a stochastic process where any finite sub-set forms a multivariate Gaussian distribution):
\begin{equation}
\label{eq:GP}
P^\mathrm{1D}(\vec{\theta}) \sim \mathcal{N}(0, K(\vec{\theta}, \vec{\theta'}; \vec{\psi})).
\end{equation}
Here, we have dropped the dependence on \(k_{||}\) and \(z\); each \(z\) bin is emulated separately and correlations between \(k_{||}\) bins are not modelled. The zero mean condition is approximated by normalising the flux power spectra by the median value in the training set. It is then necessary to specify a form for the covariance \(K(\vec{\theta}, \vec{\theta'}; \vec{\psi})\) between two points in parameter space \(\vec{\theta}\) and \(\vec{\theta'}\); note however that this is a much more general specification than in traditional interpolation methods where the functional form for \(P^\mathrm{1D}(\vec{\theta})\) must be given. We use a linear combination of a squared exponential (or radial basis function; RBF) and a linear kernel: \(K(\vec{\theta}, \vec{\theta'}; \vec{\psi}) = \sigma^2_\mathrm{RBF} \exp{\left(- \frac{(\vec{\theta} - \vec{\theta'})^2}{2 l^2}\right)} + \sigma^2_\mathrm{linear} \vec{\theta}.\vec{\theta'}\). This gives three hyper-parameters \(\vec{\psi}\): two variances for the squared exponential (\(\sigma^2_\mathrm{RBF}\)) and the linear (\(\sigma^2_\mathrm{linear}\)) kernels and a length-scale \(l\) for the squared exponential. We optimise these hyper-parameters (or ``train'' the emulator) by maximising the (Gaussian) marginal likelihood of the training data.

Once the emulator has been trained, we can use the Gaussian process to interpolate \(P^\mathrm{1D}(\vec{\theta})\) at arbitrary values of \(\vec{\theta}\). For this, we calculate the posterior predictive distribution of simulation output \(P^\mathrm{1D}(\vec{\theta^*})\) conditional on the training data \(P^\mathrm{1D}(\vec{\theta}_i)\), for \(i\) in \(\{1, \dots, N\}\), where \(N\) is the number of training samples:
\begin{equation}
\label{eq:prediction}
p(P^\mathrm{1D}(\vec{\theta^*}) | P^\mathrm{1D}(\vec{\theta}_i), \vec{\theta}_i, \vec{\theta^*}) \sim \mathcal{N}(K_* K^{-1} P^\mathrm{1D}(\vec{\theta}_i), K_{**} - K_* K^{-1} K_*^\mathrm{T}).
\end{equation}
Here, \(K_* = K(\vec{\theta^*}, \vec{\theta}_i; \vec{\psi})\) and \(K_{**} = K(\vec{\theta^*}, \vec{\theta^*}; \vec{\psi})\). Since we have determined the full distribution of interpolated flux power spectra, we have a robust estimate of the error in our interpolation as given by the variance term in eq.~\eqref{eq:prediction}. Note that this variance is independent of the training output, being only dependent on the locations in parameter space and number of training samples; this will be of use in section~\ref{sec:batch}.

\subsubsection{Likelihood function and Markov chain Monte Carlo sampling}
\label{sec:likelihood}
We now have all the pieces to construct the likelihood function and perform inference. We use a simple Gaussian likelihood function\footnote{In principle, a non-Gaussian likelihood function could be used if necessary as long as a robust statistical model can be constructed which propagates uncertainty from the emulator.}. The mean is inferred by our trained Gaussian process \(\vec{\mu}(\vec{\theta^*}) = K_* K^{-1} P^\mathrm{1D}(\vec{\theta}_i)\) [eq.~\eqref{eq:prediction}]. The covariance matrix adds in quadrature the ``data'' covariance matrix as estimated by BOSS DR9 for their real data and the diagonal ``emulator'' covariance matrix, using the variance as inferred by the trained Gaussian process \(\vec{\sigma}^2(\vec{\theta^*}) = K_{**} - K_* K^{-1} K_*^\mathrm{T}\) [eq.~\eqref{eq:prediction}]. In this way, the statistical model accounts for the uncertainty in the theoretical predictions (emulation) and we are able to test our method using realistic BOSS errors. Note that the Gaussian process is not currently modelling correlations between different scale bins. In our companion paper \cite{emulator_forest_GP_2018}, these are conservatively estimated as being maximally correlated (though uncorrelated across redshifts). Both approaches amount to approximations in the likelihood function; in future work, explicit modelling of these correlations by the Gaussian process can be investigated. These likelihood approximations do not change our main results on demonstrating the effect of Bayesian emulator optimisation for a given likelihood function. Having constructed the likelihood function, we then estimate the posterior probability distribution for our mock data \(\vec{d}\) by MCMC sampling using the \texttt{emcee} package \cite{2013PASP..125..306F}.

\subsection{Bayesian optimisation}
\label{sec:optimisation}
For Gaussian process emulators (as with machine learning problems generically), the construction of the training dataset is critical. In section~\ref{sec:training}, we detailed the use of Latin hypercube sampling to construct the training dataset. Latin hypercubes have good space-filling properties ensuring that the prior volume is well sampled. However, this is not necessarily the most efficient use of the limited resources available. In optimisation problems, such as the estimation of a posterior probability distribution where we are only interested in the peak and surrounding credible region of the distribution, evenly sampling the full prior volume may waste training samples on the edges of that volume. The idea of Bayesian optimisation is to build up the training dataset actively and iteratively using, at each stage, the information we have gained from previous iterations of the emulator. This is expressed in the acquisition function (section~\ref{sec:acquisition}). Bayesian optimisation proceeds by iteratively proposing new training data points at the maximum of the acquisition function (plus a small random displacement; section~\ref{sec:serial}), which is continually updated as the emulator is re-trained on the expanded training dataset. This procedure can be modified to allow ``batches'' of training data to be acquired simultaneously (section~\ref{sec:batch}), which may be preferred depending on the availability of computational resources.

\subsubsection{Acquisition function}
\label{sec:acquisition}
At each iteration, the next training simulation is run at the parameters of the maximum of the acquisition function (plus a small random displacement; see below). This function should peak where uncertainty in the emulated function is high so that it is better characterised. It should also peak where the objective function is high so that training samples accumulate where it matters. This manifests in a function which increases with the Gaussian process uncertainty or variance (exploration) and with the objective function itself (exploitation).

We use the Gaussian process upper confidence bound (GP-UCB) acquisition function \cite{Cox97sdo:a, auer2002using, auer2002finite, dani2008stochastic}. This is a weighted linear combination of exploitation and exploration terms, usually \(\vec{\mu}(\vec{\theta}) + \alpha \vec{\sigma}(\vec{\theta})\), where \(\vec{\mu}(\vec{\theta})\) and \(\vec{\sigma}(\vec{\theta})\) are respectively the mean and standard deviation of the posterior predictive distribution as given by the Gaussian process (see eq.~\eqref{eq:prediction}). Here, \(\vec{\theta}\) is the parameter vector and \(\alpha\) is a hyper-parameter, which can be optimised to give minimum regret\footnote{\Ie \(\mathrm{lim}_{T \rightarrow \infty} \frac{R_T}{T} = 0\), where the cumulative regret \(R_T = \sum_{t=1}^T [f_\mathrm{max} - f(\vec{\theta}_t)]\), where \(f_\mathrm{max}\) is the true maximum of the objective function \(f\) and \(\vec{\theta}_t\) are the positions of acquisition function proposals for the training set. Minimising regret puts a lower limit on the convergence rate of finding the true optimum.} in the decisions made (ref.~\cite{srinivas2009gaussian} calculate optimal values for various Gaussian process covariance kernels when the objective is the emulated function). A subtlety arises because in this study, we are not emulating the function which must be optimised (the posterior), but rather the flux power spectrum, which, along with its uncertainty, goes into the likelihood function (see section~\ref{sec:likelihood}). We therefore use a modified form of the acquisition function:
\begin{equation}
\label{eq:acquisition}
\mathcal{A}(\vec{\widetilde{\theta}}) = \mathcal{P}(\vec{\widetilde{\theta}} | \vec{d}) + \alpha \vec{\sigma}^\mathrm{T}(\vec{\widetilde{\theta}}) \Sigma^{-1} \vec{\sigma}(\vec{\widetilde{\theta}}),
\end{equation}
where \(\mathcal{P}(\vec{\widetilde{\theta}} | \vec{d})\) is the logarithm of the posterior probability distribution given data \(\vec{d}\) and \(\Sigma\) is the data covariance matrix (see section~\ref{sec:likelihood}). This function exploits the current best estimate of the posterior (by the first term) but explores the parameter space where this estimate is most uncertain (by the second term). The second term is constructed by estimating the uncertainty on the (log) posterior as \(\frac{1}{2} |\mathcal{P}(\vec{\mu}(\vec{\widetilde{\theta}}) + \vec{\sigma}(\vec{\widetilde{\theta}})) - \mathcal{P}(\vec{\mu}(\vec{\widetilde{\theta}}) - \vec{\sigma}(\vec{\widetilde{\theta}}))|\).

A further subtlety is that although the mean flux parameters \([\tau_0, \mathrm{d}\tau_0]\) are emulated, an arbitrary number of training samples can be constructed for these parameters since their effect on the flux power spectrum is estimated in the post-processing of hydrodynamical simulations (section~\ref{sec:training}). Therefore, no Bayesian optimisation is necessary for the training set in these dimensions. It follows that the posterior in eq.~\eqref{eq:acquisition} is marginalised over the mean flux parameters and that \(\vec{\widetilde{\theta}}\) only lists the remaining cosmological and astrophysical parameters (see section~\ref{sec:training}). For the hyperparameter \(\alpha\), we use the optimised form as given by refs.~\cite{srinivas2009gaussian, 2010arXiv1012.2599B}. Here, we simplify their notation to \(\alpha = 0.97 \sqrt{\nu}\) and \(\nu\) linearly decreases from a starting value of 1 to 0.4 by convergence as the size of the training set increases in order to reflect the increasing confidence in the emulator. Since \(\nu \sim 1\), this approximates to about a \(1 \sigma\) uncertainty in the log posterior.

Gaussian processes are unsuited to extrapolation and so the emulator error increases sharply at the edges of the prior volume. This can spuriously dominate the acquisition function (we manifestly do not want to add samples on the perimeter of the prior volume). Therefore, we apply a uniform prior when finding the maximum of the acquisition function which excludes the outer 5 \% of parameter space in each dimension (approximating the convex hull formed by the initial training samples).

Finally, following \eg refs.~\cite{2015arXiv150103291G, 2018PhRvD..98f3511L}, once the maximum of the acquisition function has been found, the final proposal for a new training sample is a small random displacement away from the maximum. This helps to ensure that the same position in parameter space cannot be proposed more than once, which can in principle be the case especially when the emulator is (near-)converged. It also helps to explore the credible region around the peak of the posterior distribution (to the extent desired by the user, usually the 95\% credible region), bearing in mind that the optimisation procedure should not only find the peak of the distribution but should correctly characterise the region around it. Therefore, this random displacement can be drawn from a Gaussian distribution with a full-width-at-half-maximum set by the current estimation of the posterior contours (most simply approximated by using the desired number (\eg two) of sigma from the 1D marginalised distributions; see ref.~\cite{2017arXiv170400520J} for discussion and comparison of deterministic and stochastic acquisition rules). In general, these ``exploration'' terms should be tuned to the size of the credible region which is required to be well estimated.

\subsubsection{Initial Latin hypercube}
\label{sec:hypercube}
We initiate the Bayesian optimisation with a Latin hypercube (see section~\ref{sec:training}; other sampling schemes with good space-filling properties, \eg a Sobol sequence \cite{sobol1967distribution}, can be used) on the full prior volume. The size of this initial hypercube should not be so small as to characterise the emulated function poorly. As mentioned in section~\ref{sec:training}, when the density of training samples (for the Latin hypercube) in the prior volume \(N_\mathrm{Latin} / V_\mathrm{prior}\) is too low, the emulated function is characterised so poorly that the likelihood function, in propagating emulator error, is biased. In testing with an initial hypercube of only nine simulations, the convergence rate of the Bayesian optimisation from this initialisation became such that it was more efficient to run an initial hypercube with more simulations. Equally, the size of the initial hypercube should not be too large. This would waste samples on the edges of the prior volume and negate the power of Bayesian optimisation to propose training samples efficiently. Ultimately, the best size for the initial hypercube is a trade-off and we show the results of our experiments (our initial hypercube has 21 simulations) in section~\ref{sec:results}, with discussion in section~\ref{sec:discussion}.

\subsubsection{Serial optimisation}
\label{sec:serial}
Serial Bayesian optimisation proceeds by proposing and running training simulations one-by-one. After each training simulation has been added to the training dataset, the emulator is re-trained and the acquisition function re-evaluated. The procedure can be summarised by these steps:
\begin{enumerate} 
\item \label{item:initial}Construct the initial training dataset by running hydrodynamical simulations (see section~\ref{sec:training} for details) at parameters sampled by a Latin hypercube on the full prior volume (section~\ref{sec:hypercube}). (Also see our companion paper \cite{emulator_forest_GP_2018} for more discussion about using Latin hypercubes.)
\item \label{item:GP}Train the Gaussian process emulator for the flux power spectrum (section~\ref{sec:interpolation}) and then evaluate the posterior probability distribution for the given data (section~\ref{sec:likelihood}).
\item \label{item:acquisition}Evaluate the acquisition function (eq.~\eqref{eq:acquisition}) and find its maximum in order to propose the location of the next training simulation (plus a small random displacement; section~\ref{sec:acquisition}).
\item \label{item:refinement}Run this ``refinement'' simulation, re-train the emulator using the optimised and expanded training set and then re-evaluate the posterior distribution.
\item \label{item:iteration}Repeat the previous \emph{two} steps (\ref{item:acquisition} and \ref{item:refinement}) until the desired number of optimisation steps have been executed.
\item \label{item:convergence}Optimisation can continue until successive estimations of the posterior (practically, summary statistics like the mean and \(1 \sigma\) limits can be used) are seen to sufficiently converge (to some specified tolerance like a fraction, \eg 0.2, of a sigma).
\item \label{item:validation} Performance can be checked by cross-validation and/or a test suite of fiducial simulations (bearing in mind that the emulator is optimised only for the true parameters of the data but that tests with fiducial simulations will still inform us about performance in relevant areas of parameter space).
\end{enumerate} 

\begin{figure}[tbp]
\centering 
\includegraphics[width=\textwidth]{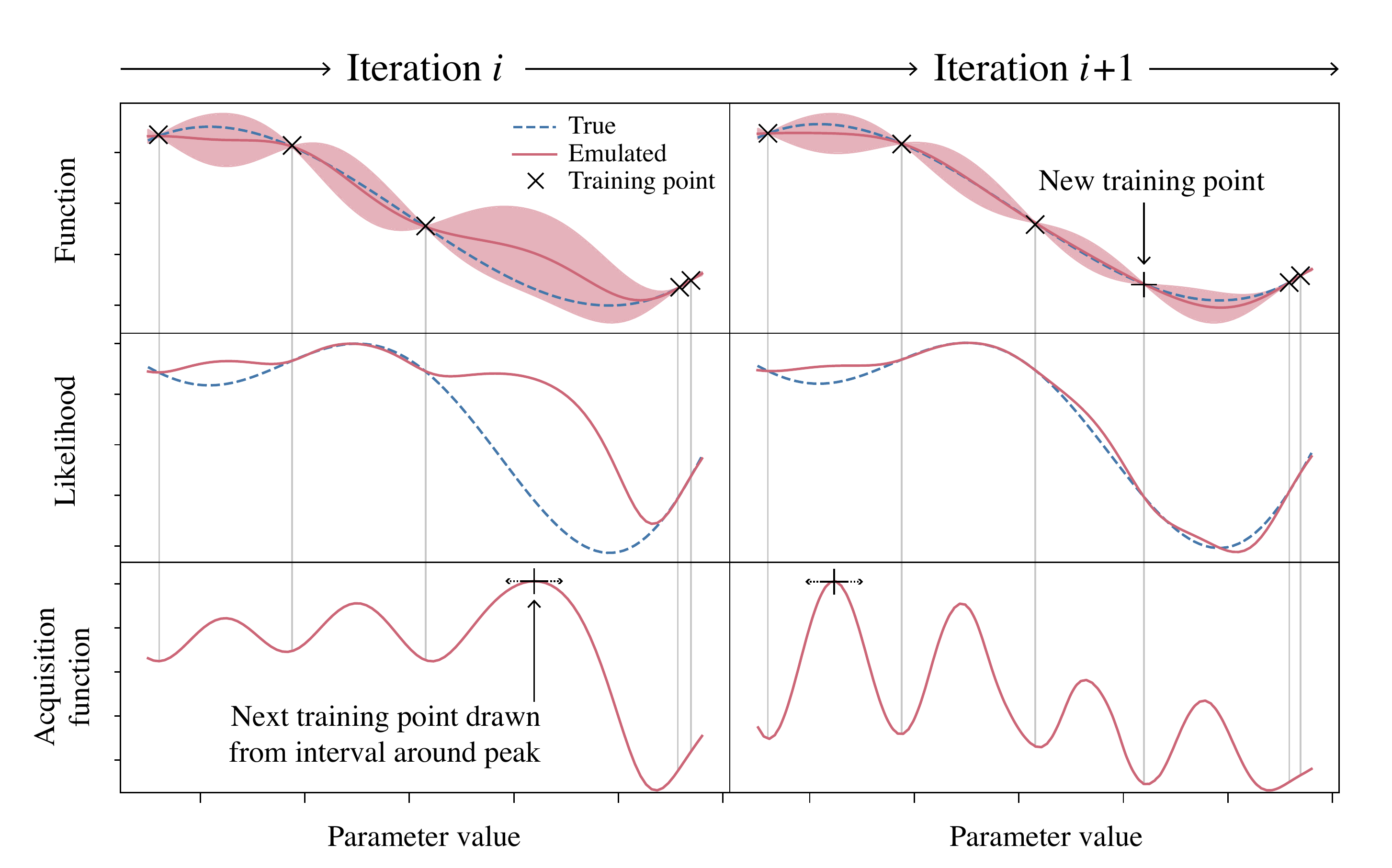} 
\caption{\label{fig:optimisation}An illustration of (\emph{from left to right}) two example successive iterations of the serial Bayesian optimisation (section~\ref{sec:serial}) for a toy function. \emph{From top to bottom}, we show the emulated function (the truth is a cubic function); the (Gaussian) likelihood function; and the acquisition function. The blue dotted lines show the truth and the red solid lines show the emulated estimation of the toy function (the mean as inferred by the Gaussian process model; see section~\ref{sec:interpolation}) in the top panels and the likelihood propagating the emulator error (as described in section~\ref{sec:likelihood}) in the middle panels. The red band shows the \(\pm 1 \sigma\) error on the emulated function as inferred by the Gaussian process. The training data (with which the Gaussian process is optimised) are indicated by the black crosses. At the end of each iteration, the next training sample is proposed at the maximum of the acquisition function (plus a small random displacement; see section~\ref{sec:acquisition}). Bayesian optimisation efficiently proposes new training samples in order to better characterise the emulated function and therefore the true likelihood function.}
\end{figure}

Figure~\ref{fig:optimisation} illustrates two example iterations of the serial optimisation method. It shows the procedure for a toy (cubic) function. It demonstrates how Bayesian optimisation efficiently proposes training samples in order to better characterise the emulated function and therefore the true likelihood function. In the actual case, the emulated function is the 1D flux power spectrum and there are six model parameters (three cosmological, two astrophysical and the mean flux at each redshift; see section~\ref{sec:training}).

\subsubsection{Batch optimisation}
\label{sec:batch}
Batch Bayesian optimisation proceeds as in the serial case (section~\ref{sec:serial}) except that at each optimisation step, multiple training samples are proposed and evaluated simultaneously in a single batch. This may be preferred depending on the particular allocation of computational resources for running hydrodynamical simulations. Within each batch, decisions are in fact still made in series but without running the simulations from earlier decisions or re-training the emulator until the set of batch proposals is completed. The positions in parameter space of new proposals from earlier on in the batch can be added to the training set in order to help inform later decisions within the same batch. This will correctly update the expected error distribution of the emulator. We can do this because the variance of a Gaussian process is independent of the training output (it is only a function of the Gaussian process covariance kernel; see section~\ref{sec:interpolation} and eq.~\eqref{eq:prediction}). Thus, in turn, before each proposal in a given batch, the second ``exploration'' term in the acquisition function (eq.~\eqref{eq:acquisition}) can be updated and a new maximum found. The disadvantage is that, even so, proposals later on in a given batch are less informed about the true objective function (for us, the posterior distribution) than they would be in the equivalent serial case.

Once all the proposals in a given batch have been made and the simulations have finished running, the emulator can be re-trained and the acquisition function fully re-evaluated. A new batch can be started as necessary. For maximum efficiency of the Bayesian optimisation, the procedure should be as serial as possible so that each proposal is as informed as it can be; \ie the batch size should be as small as the allocation of computational resources reasonably allows. In our testing in section~\ref{sec:results}, we use a batch size of three. The procedure can be summarised by these steps:
\begin{enumerate} 
\item \label{item:initial}As with serial optimisation (section~\ref{sec:serial}), the procedure starts with an initial Latin hypercube training set, which is used to train a first iteration of the emulator (see section~\ref{sec:hypercube}).
\item \label{item:first}The first proposal of each optimisation batch is made using the maximum of the full acquisition function as given by eq.~\eqref{eq:acquisition} (see section~\ref{sec:acquisition}).
\item \label{item:batch}Subsequent proposals in each batch use the acquisition function, partially updated with the new Gaussian process standard deviation \(\vec{\sigma}(\vec{\theta})\) including the effect of proposals made previously in the same batch. This is achieved by adding to the training set the positions in parameter space of earlier proposals from the same batch without re-training the emulator. This exploits the fact that the variance of a Gaussian process is independent of training output.
\item \label{item:iteration}Within each batch, repeat the above step (\ref{item:batch}) until the desired number of samples per batch has been chosen.
\item \label{item:parallel}Simultaneously run all the ``refinement'' simulations of the batch, re-train the emulator using the optimised and expanded training set (including its simulation output) and then re-evaluate the posterior distribution.
\item \label{item:batch_iteration}Repeat the previous \emph{four} steps (\ref{item:first} to \ref{item:parallel}) until the desired number of optimisation batches have been executed.
\item \label{item:convergence}The same convergence and cross-validation tests can be carried out as in the serial case (see section~\ref{sec:serial}).
\end{enumerate} 

\section{Results}
\label{sec:results}

\subsection{Serial optimisation}
\label{sec:serial_results}

\begin{figure}[tbp]
\centering 
\includegraphics[width=\textwidth]{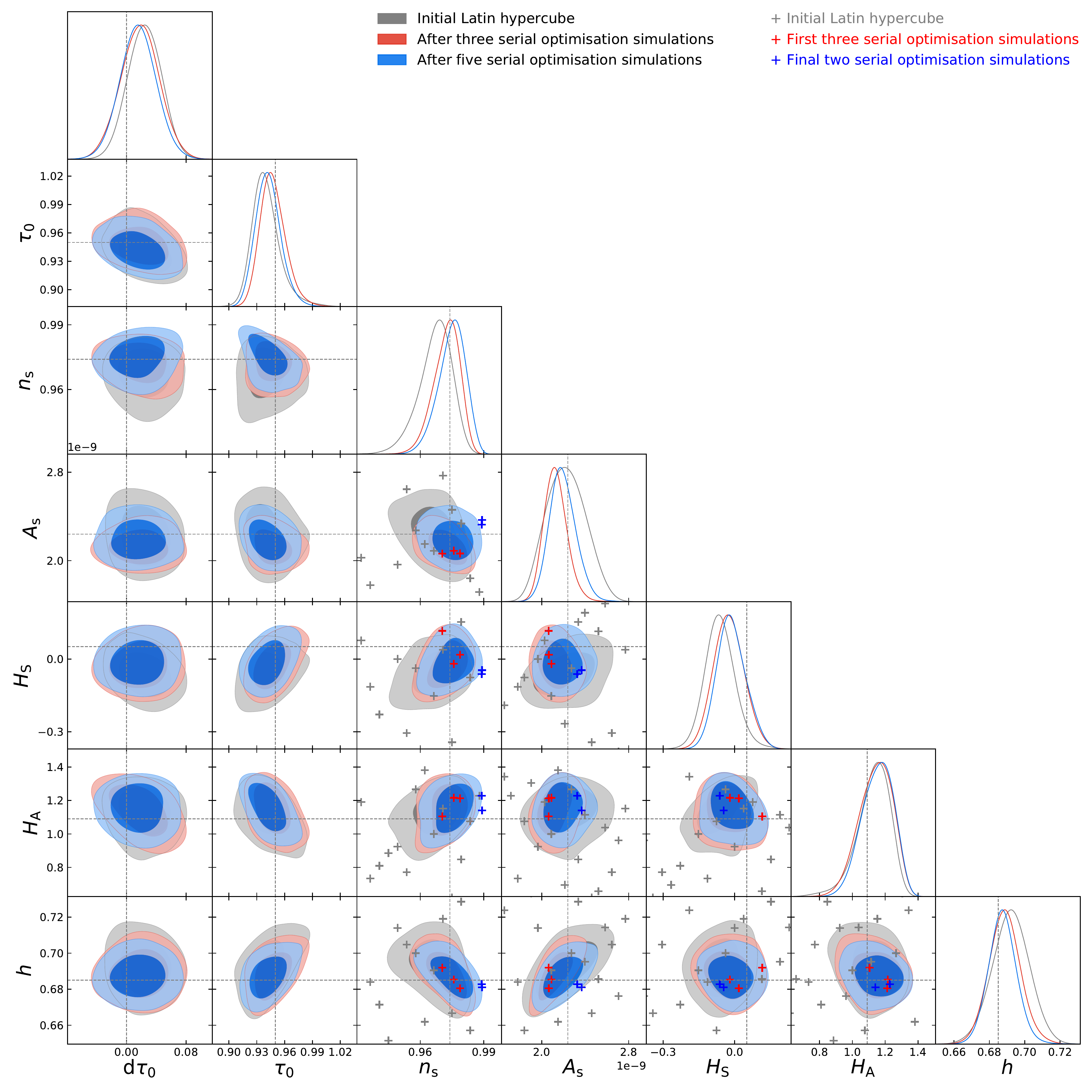} 
\caption{\label{fig:serial}The 1D and 2D marginalised posterior probability distributions (see section~\ref{sec:likelihood}) for the seven model parameters \(\vec{\theta}\) (section~\ref{sec:training}) given the mock data vector \(\vec{d}\). The gray, red and blue contours use an emulator trained respectively on the initial Latin hypercube of 21 samples; the initial hypercube plus three serial Bayesian optimisation simulations; and the initial hypercube plus five serial optimisation simulations. The darker and lighter shaded contours show respectively the 68\% and 95\% credible regions. The gray dotted lines indicate the true model parameter values \(\vec{\theta_\mathrm{true}}\) of our mock data vector. The gray, red and blue crosses indicate the projected positions in parameter space of our training samples respectively for the initial Latin hypercube; the first three serial optimisation simulations; and the final two serial optimisation simulations. Note that we do not show the training samples for the mean flux parameters \([\mathrm{d}\tau_0, \tau_0]\), which do not form part of our Latin hypercube or Bayesian optimisation and are treated differently (see section~\ref{sec:training} for details).}
\end{figure}

Figure~\ref{fig:serial} shows the results of serial Bayesian optimisation (section~\ref{sec:serial}) on the (1D and 2D marginalised) posterior probability distribution (in the filled coloured contours) of our model parameters \(\vec{\theta}\) given our mock data \(\vec{d}\) (the true parameters \(\vec{\theta_\mathrm{true}}\) of which are indicated by the gray dotted lines)\footnote{Note that, in general, our inferred posterior distributions differ from our companion paper \cite{emulator_forest_GP_2018} in their width due to different approximations in the likelihood function (see section~\ref{sec:likelihood}). However, this does not change our main conclusions on the effect of Bayesian emulator optimisation.}. The coloured crosses indicate the projected positions in parameter space of our training simulations. Note that the mean flux parameters \([\mathrm{d}\tau_0, \tau_0]\) do not form part of the Latin hypercube and are treated differently (see section~\ref{sec:training} for more details). Most importantly, these parameters are adjusted in the post-processing of hydrodynamical simulations and hence it is not necessary to employ Bayesian optimisation in these axes. Otherwise, our initial Latin hypercube (section~\ref{sec:hypercube}; in gray; consisting of 21 simulations) fills the full prior volume with random samples, uniformly in projection on each parameter axis. By employing the procedure detailed in section~\ref{sec:serial}, five optimisation samples were chosen (until the inferred posterior distributions were seen to converge as confirmed by subsequent optimisation steps by the tests and details set out in section~\ref{sec:serial}). The first three of these are coloured in red and the final two in blue. Thanks to the Bayesian optimisation exploiting our knowledge of the approximate posterior distribution as inferred using previous iterations of the emulator, these optimisation samples are concentrated in the most important region of parameter space (the 95\% credible region of the posterior distribution). This region is explored by including the emulator error \(\vec{\sigma}(\vec{\theta})\) in the Bayesian optimisation acquisition function (eq.~\eqref{eq:acquisition}) and the stochastic element in the final acquisition (see section~\ref{sec:acquisition}). Note that although the projection makes some of the hypercube samples seem to appear in the peak of the posterior, it is the optimisation samples that are actually located in the central posterior volume.

It can be seen in figure~\ref{fig:serial} that reduced emulator error after Bayesian optimisation propagates to reducing the widths of the marginalised posterior distributions, \eg the \(1 \sigma\) error on \(A_\mathrm{s}\) reduces by 38\%. The effect of the Bayesian optimisation is to reduce the emulator error in the central posterior volume. For example, the emulator error at \(\vec{\theta_\mathrm{true}}\) is reduced by 61\% (averaged over all the \(k_{||}\) and \(z\) bins of the data vector) after the five serial optimisation steps; meaning that the ratio of emulator error to ``data'' error (the diagonal elements of the data covariance matrix; see section~\ref{sec:likelihood}), again averaged over all bins, goes from 16\% for the initial hypercube to 6\% after optimisation. By reducing this ratio, the more accurate emulator provides a more accurate estimation of the posterior distribution (see eq.~\eqref{eq:acquisition} and section~\ref{sec:acquisition}). Figure~\ref{fig:optimisation} shows schematically how this effect works. In general, when the error in the emulated function is high, the likelihood at those parameter values, in propagating the error, is inflated. This is because the increased uncertainty means that there is more probability than otherwise that the data are drawn from that region of parameter space. The general effect is to broaden the peak of the likelihood function. However, the addition of new training points better characterises the interpolated function and therefore the true likelihood. (Spurious biases, rather than simple broadening of the peak, can also arise in the limit of the emulator-to-data error ratio being too large, as is discussed in section~\ref{sec:initial_hypercube_discussion}.)

\subsection{Batch optimisation}
\label{sec:batch_results}

\begin{figure}[tbp]
\centering 
\includegraphics[width=\textwidth]{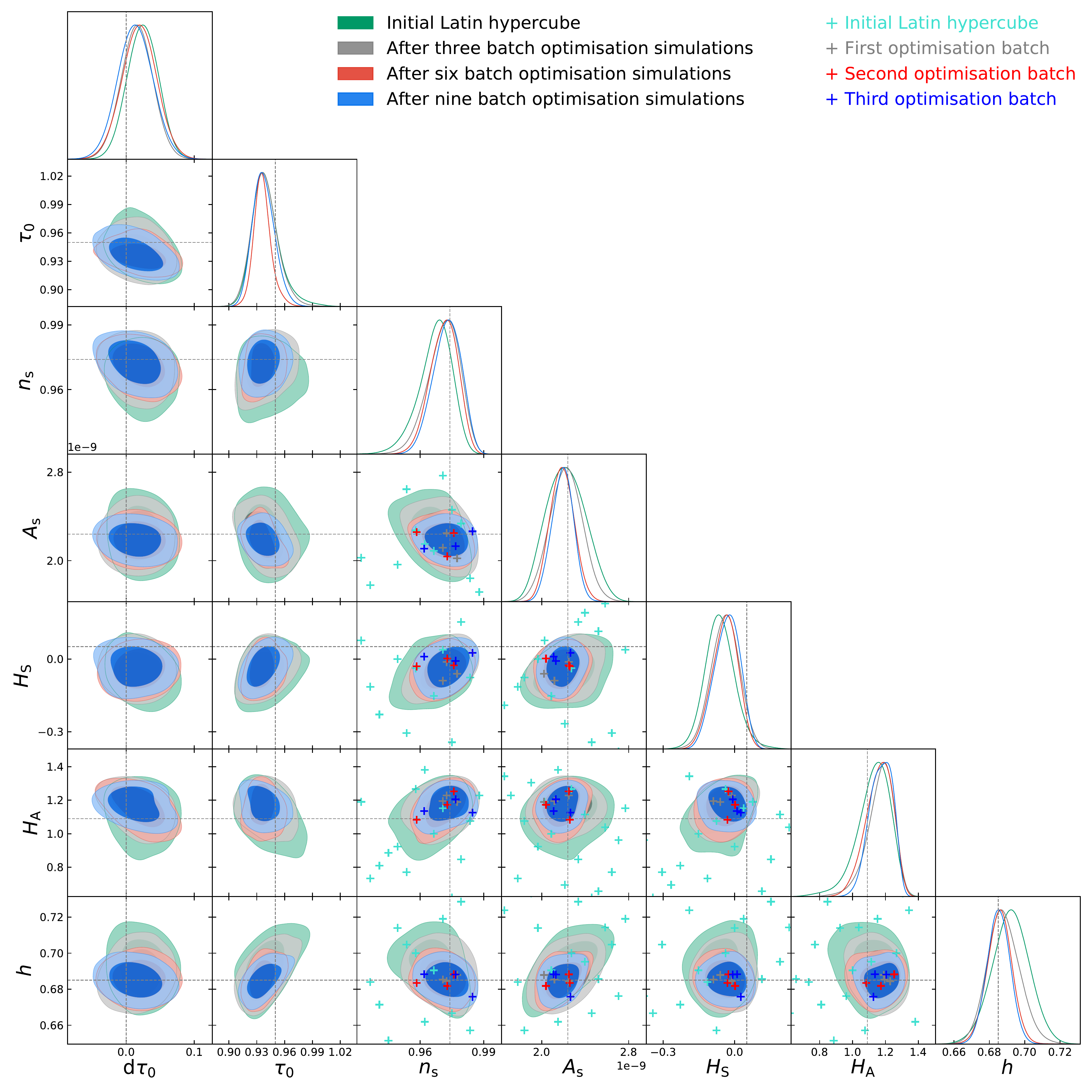} 
\caption{\label{fig:batch}As figure~\ref{fig:serial} except showing the results of batch Bayesian optimisation.}
\end{figure}

Figure~\ref{fig:batch} shows the results of batch Bayesian optimisation (section~\ref{sec:batch}) on the posterior probability distribution of our model parameters \(\vec{\theta}\) given our mock data \(\vec{d}\); again, the coloured crosses indicate the projected positions in parameter space of our training simulations. The difference from above is that now optimisation samples are proposed in batches of three. The results and explanations are very similar to the serial case (see above): the optimisation samples are concentrated in the central posterior volume exploiting our knowledge of the posterior; the emulator-to-data error ratio is reduced to the same extent; and the widths of the marginalised posterior distributions shrink. The main difference is that the batch optimisation takes longer to converge (after three batches each of three simulations). This makes sense since the second and third proposals in each batch are less informed than the equivalent serial proposal because the emulator is only re-trained after each batch of three has finished running in parallel. We will discuss the benefits and disadvantages of batch acquisition in section~\ref{sec:discussion}.

\subsection{Comparison to a Latin hypercube}
\label{sec:comparison_results}

\begin{figure}[tbp]
\centering 
\includegraphics[width=\textwidth]{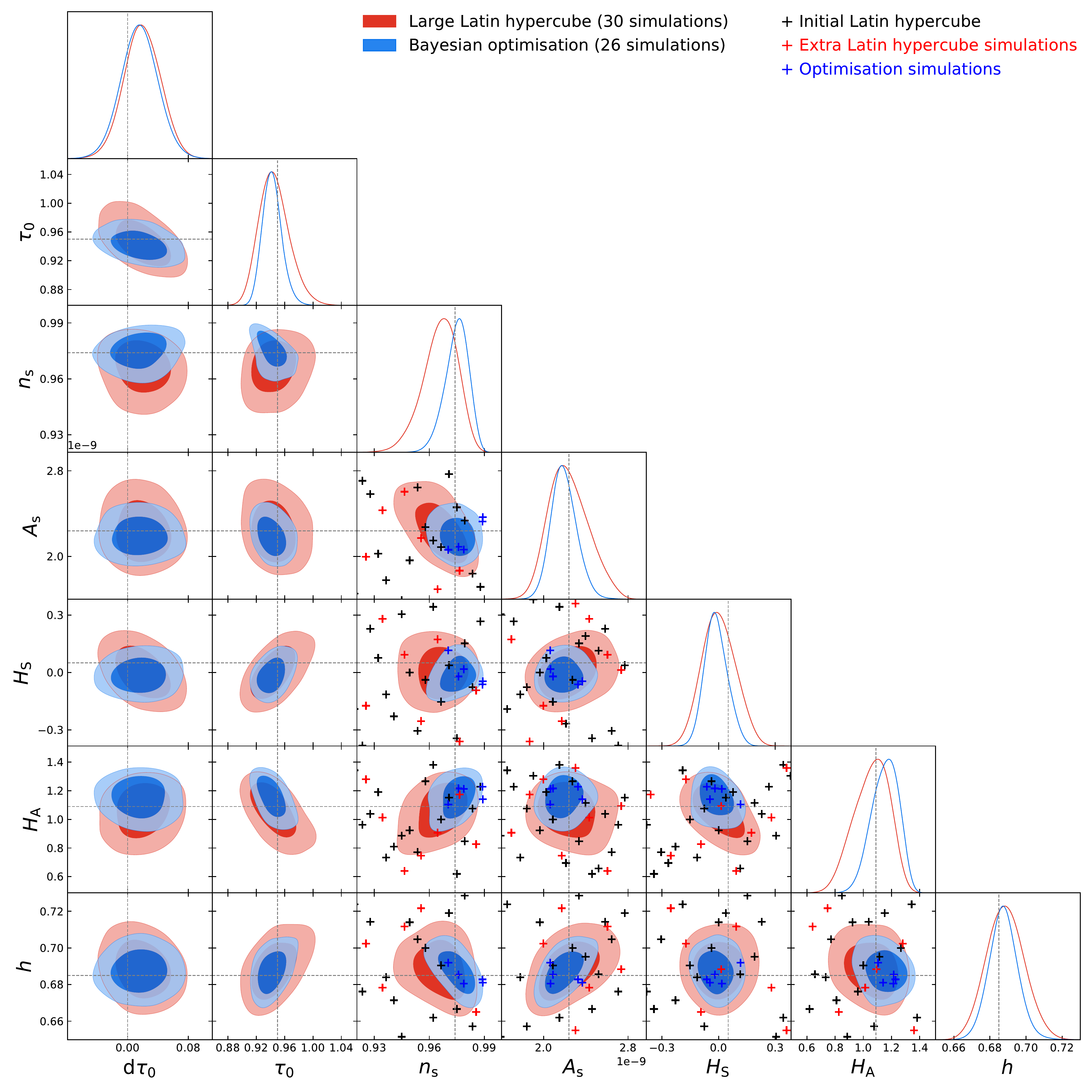} 
\caption{\label{fig:optimisation_latin}As figure~\ref{fig:serial} except comparing the results of Bayesian optimisation (with a total number of 26 training simulations) to a Latin hypercube (with 30 training simulations).}
\end{figure}

Figure~\ref{fig:optimisation_latin} compares the results of Bayesian optimisation with a Latin hypercube of 30 simulations (our initial hypercube from above from which we optimise has 21 simulations). For a fair comparison, we construct this larger Latin hypercube (still spanning the full prior volume) using the initial Latin hypercube as a sub-set of its samples; the extra nine training points are indicated by the red crosses in figure~\ref{fig:optimisation_latin}. It can be seen in figure~\ref{fig:optimisation_latin} that the smaller emulator error from Bayesian optimisation with respect to the Latin hypercube propagates to reducing the size of the 68\% and 95\%  credible regions of the posterior distribution. Indeed, the full volume of these regions reduces by 90\% and, \eg the \(1 \sigma\) error on \(A_\mathrm{s}\) reduces by 38\%. Although the Latin hypercube has a larger training set than our serial Bayesian optimisation example (which has 26 simulations in total), because the samples of the larger hypercube are spread throughout the prior space, the Latin hypercube actually has larger emulator error in the central posterior volume. The large Latin hypercube has \(\sim 20 \%\) greater emulator error at \(\vec{\theta_\mathrm{true}}\). A more accurate emulation of the flux power spectrum means more accurate estimation of the posterior probability distribution (see above, section~\ref{sec:discussion} and figure~\ref{fig:optimisation} for explanation). This in particular means that the weakening of parameter constraints from uncertainty in forward modelling (emulation) can be reduced.

\section{Discussion}
\label{sec:discussion}

Figures~\ref{fig:serial} and \ref{fig:batch} demonstrate the ability of Bayesian optimisation to determine more accurately the posterior distribution in the central, high-probability parameter space. It achieves this by concentrating training samples for the Gaussian process emulator of the 1D Lyman-alpha forest flux power spectrum (section~\ref{sec:gaussian_process}) in the parameter space corresponding to the central, high-probability region of the posterior distribution. As a consequence, the error on the emulated flux power spectrum is more than halved (\eg a 61\% reduction when averaged over all power spectrum bins at \(\vec{\theta_\mathrm{true}}\), the true model parameters of our mock data vector). More accurate estimation of the flux power spectrum means more accurate estimation of the posterior probability distribution. This is because, in the likelihood function (see section~\ref{sec:likelihood}), the theory prediction has smaller interpolation error and in the covariance matrix, the ratio of emulator-to-data error is reduced. Figure \ref{fig:optimisation} demonstrates how this more accurate determination of the ``true'' (zero emulator error) likelihood in general manifests in reducing the width of the peak. This is because areas of parameter space with high (interpolation) uncertainty in the flux power spectrum have increased possibility than otherwise that the data are drawn from there.

We also consider how well the Gaussian process estimates interpolation error. There is detailed discussion on this matter in our companion paper \cite{emulator_forest_GP_2018}, where the estimated error distribution is compared to the true error distribution as evaluated at a suite of test simulations. The general tendency is for the Gaussian process to overestimate the emulator error moderately. This will tend to have a conservative effect on parameter estimation as it will tend to broaden the peak of the posterior distribution more than necessary.

\subsection{Comparison between serial and batch acquisition}
\label{sec:serial_batch_discussion}

The difference between the two cases in figures~\ref{fig:serial} and \ref{fig:batch} is in respectively the serial (section~\ref{sec:serial}) and batch (section~\ref{sec:batch}) acquisition. The batch optimisation takes longer to converge in its posterior distribution (requiring four more simulations). This is because training sample proposals are less informed than in the equivalent serial case (after the first proposal in each batch). The scaling of our simulation code with the number of computing cores means that, in this case, the serial acquisition was marginally more efficient in terms of overall computing time. However, we have not yet carried out detailed tests of how the convergence rate scales with batch size. Therefore, in future uses of this method, it will probably still be preferable to use batch acquisition in order to make efficient use of the available computational resources. The size of the batch to use can be determined by balancing the specifics of the distribution of computational power, the optimal load balancing of the forward modelling code and the decrease in the convergence rate from batch acquisition. A potential approach is to start with serial optimisation until the estimation of posterior distributions for the validation set is unbiased (\eg the truth is recovered to 68\% credibility) and then to continue with batch optimisation once exploration of the 95\% credible region is required.

\subsection{The importance of the initial sampling density}
\label{sec:initial_hypercube_discussion}

In our examples, the initial Latin hypercube (section~\ref{sec:hypercube}), from which we optimise, has 21 training simulations. We found this to be a sufficient number to give a reasonable first estimate of the posterior distribution. We found in testing with an initial hypercube of nine simulations (with emulator error on the same order of magnitude as the ``measurement'' error) that using fewer simulations in the initial hypercube (or, more specifically, having a lower density of initial training samples in the prior volume \(N_\mathrm{Latin} / V_\mathrm{prior}\)) can lead to significant bias and/or spurious multi-modality in the inferred posterior distribution using this initial training set. (In particular, using Bayesian optimisation from this initial set proved less efficient than simply starting from a hypercube with more simulations and a better characterisation of the objective function -- the likelihood.) Figure~\ref{fig:optimisation} helps to qualitatively understand this phenomenon. Where the emulator error is high, the likelihood function is inflated taking account of the uncertainty in model prediction. If the error is sufficiently high, it can form local maxima (\ie multi-modality) in the likelihood function, which are not otherwise present in the limit of zero emulator error.

However, the precise threshold on \(N_\mathrm{Latin} / V_\mathrm{prior}\) to keep the bias at a level below \(2 \sigma\) (by reducing the ratio of emulator to data error) has not been determined. It is additionally dependent on the particular correlation structure of the model parameter space (as estimated by the optimal Gaussian process hyper-parameter values; see section~\ref{sec:interpolation}), \ie the extent to which any particular training sample can predict model values elsewhere in the prior volume. In particular, each correlation length of the parameter volume should be sampled at least once. It follows that this is a non-trivial, problem-dependent decision. It is possible to construct Latin hypercubes with more simulations using an existing Latin hypercube as a subset of the samples, as we demonstrate in section~\ref{sec:results}. This problem of determining the size of the initial Latin hypercube could be tackled by iteratively increasing its size until the spurious biases are removed. In future uses of this method with more precise datasets, in order to maintain the same level of final emulator-to-data error ratio (\(\sim 10 \%\)), it will be necessary to reduce the emulator error further. This may require a higher density of training samples, especially in the central posterior volume using Bayesian optimisation, but also in the initial training set (in order to avoid the spurious biases discussed above). Future testing of the method with alternative model parameterisations and data vectors, with different optimal Gaussian process hyper-parameter values, (\eg the 3D flux power spectrum; see our companion paper \cite{emulator_forest_GP_2018} for more discussion) will address these important issues further.

\subsection{Comparison to a Latin hypercube}
\label{sec:comparison_discussion}

Figure~\ref{fig:optimisation_latin} demonstrates the power of Bayesian optimisation compared to ``brute-force'' Latin hypercube sampling. Following the explanations from above, smaller emulator error reduces the size of the 68\% and 95\% credible regions of the posterior volume by 90\%, with, \eg a 38\% reduction in the \(1 \sigma\) error on \(A_\mathrm{s}\), for the posterior using the Bayesian optimisation emulator compared to the Latin hypercube with four more simulations. Because the emulator error varies as a function of parameter value, there is no simple way to estimate how it affects the shape of the posterior distribution. The extreme limit of this with few training points and high emulator error leads to the spurious bias and multi-modality in the likelihood function discussed in section~\ref{sec:initial_hypercube_discussion} (rather than just simply broadening the posterior). Ultimately, the most robust method is the Bayesian optimisation which will concentrate training samples at the peak of the posterior. This will smoothly lower the emulator error at the peak of the posterior in order to determine more accurately the 95\% credible volume.

The details of optimal batch size and the size of the initial training set will be explored further in future work and is expected to be specific to the particular survey (data errors) and distribution of computational resources available. Our results (see also our companion paper \cite{emulator_forest_GP_2018}) on BOSS DR9 -- the current state-of-the-art in terms of large-scale survey 1D flux power spectrum -- mock data have shown that Bayesian optimisation can lead to smaller errors propagating through to the final model parameter estimation. This is achieved with fewer simulations than simply running a Latin hypercube with more simulations.

\section{Conclusions}
\label{sec:conclusions}

We have investigated Bayesian emulator optimisation -- iterative addition of extra simulation samples to an existing emulator of the Lyman-alpha forest 1D flux power spectrum. We found that this method produces converged posterior parameter constraints with 15\% fewer simulations than a single-step Latin hypercube. Bayesian optimisation of the training set reduces the error (compared to brute-force Latin hypercube sampling) in the required interpolation (emulation) of simulated flux power spectra. This propagates to more accurate inference of the posterior distribution. In our companion paper \cite{emulator_forest_GP_2018}, we showed how Gaussian processes can be used to robustly interpolate between a training set of simulations so that the likelihood function can be evaluated. Gaussian processes give a principled estimate of the error in this interpolation which can be propagated to the final inference.

However, the construction of the training set is essential, especially considering how few simulations are available. We found that building the training set using Bayesian optimisation concentrates training samples in areas of high posterior probability (\ie the 95\% credible region), where we most care about correctly inferring the posterior distribution. This is akin to the standard MCMC approaches to sampling posterior distributions, which focus on characterising the peak of the distribution accurately. Bayesian optimisation achieves this by iteratively proposing training samples, exploiting knowledge of the approximate posterior from previous iterations of the emulator and exploring the prior volume where our current characterisation of the posterior is most uncertain. We initiated the procedure with a Latin hypercube training set; we found that if this initial set has too low a density of samples in the prior volume, the emulator error is high enough to bias the inferred posterior distribution. We explored two types of acquisition for the Bayesian optimisation: one where proposals are made serially and the decisions are maximally informed; and one where proposals are made in batches and simulations are run in parallel. The latter can make more efficient use of computational resources at the expense of some increase in the number of optimisation steps until convergence.

We found that both Bayesian optimisation acquisition methods can give smaller emulator error in interpolated flux power spectra in areas of high posterior probability than by using a Latin hypercube to construct the training dataset. Despite having four more simulations in the training set, the emulator using a large Latin hypercube had \(\sim 20 \%\) larger error at the true parameters of our mock data than by using Bayesian optimisation to concentrate training samples around the peak of the posterior. This reduced emulator error propagates to reduced widths on the inferred posterior distribution, with the volume of the 68\% and 95\% credible regions shrunk by an order of magnitude and, \eg a 38\% reduction in the \(1 \sigma\) error on the amplitude of the small-scale primordial fluctuations (\(k_\mathrm{pivot} = \frac{2 \pi}{8}\,\mathrm{Mpc}^{-1}\)).

This is the first demonstration of Bayesian optimisation applied to large-scale structure emulation. Having tested the efficacy of Bayesian emulator optimisation on mock data with BOSS DR9 data covariance, this study outlines a new methodology for the cosmological and astrophysical parameter estimation from ongoing and future spectroscopic surveys like eBOSS and DESI. Furthermore, we anticipate that these methods will be of benefit to the many other emulators of the large-scale structure (\eg galaxy clustering, weak lensing and 21cm).

\acknowledgments

The authors thank Jens Jasche, Florent Leclercq, Chris Pedersen and Risa Wechsler for valuable discussions. This work was performed in part at the Aspen Center for Physics, which is supported by National Science Foundation (NSF) grant PHY-1607611. KKR was supported by the Science Research Council (VR) of Sweden. HVP was partially supported by the European Research Council (ERC) under the European Community's Seventh Framework Programme (FP7/2007-2013)/ERC grant agreement number 306478-CosmicDawn, and the research project grant ``Fundamental Physics from Cosmological Surveys'' funded by the Swedish Research Council (VR) under Dnr 2017-04212. AP was supported by the Royal Society. HVP and AP were also partially supported by a grant from the Simons Foundation. SB was supported by NSF grant AST-1817256. LV was supported by the European Union's Horizon 2020 research and innovation programme ERC (BePreSySe, grant agreement 725327) and Spanish MINECO projects AYA2014-58747-P AEI/FEDER, UE, and MDM-2014-0369 of ICCUB (Unidad de Excelencia Maria de Maeztu). AFR was supported by a Science and Technology Facilities Council (STFC) Ernest Rutherford Fellowship, grant reference ST/N003853/1. HVP, AP and AFR were further supported by STFC Consolidated Grant number ST/R000476/1. This work was partially enabled by funding from the University College London (UCL) Cosmoparticle Initiative.


\bibliographystyle{JHEP}
\bibliography{refinement}

\providecommand{\href}[2]{#2}\begingroup\raggedright\begin{thebibliography}{10}

\bibitem{2005PhRvD..71j3515S}
U.~{Seljak}, A.~{Makarov}, P.~{McDonald}, S.~F. {Anderson}, N.~A. {Bahcall},
  J.~{Brinkmann} et~al., \emph{{Cosmological parameter analysis including SDSS
  Ly{$\alpha$} forest and galaxy bias: Constraints on the primordial spectrum
  of fluctuations, neutrino mass, and dark energy}},
  \href{https://doi.org/10.1103/PhysRevD.71.103515}{\emph{\prd} {\bfseries 71}
  (2005) 103515} [\href{https://arxiv.org/abs/astro-ph/0407372}{{\ttfamily
  astro-ph/0407372}}].

\bibitem{2015JCAP...11..011P}
N.~{Palanque-Delabrouille}, C.~{Y{\`e}che}, J.~{Baur}, C.~{Magneville},
  G.~{Rossi}, J.~{Lesgourgues} et~al., \emph{{Neutrino masses and cosmology
  with Lyman-alpha forest power spectrum}},
  \href{https://doi.org/10.1088/1475-7516/2015/11/011}{\emph{\jcap} {\bfseries
  11} (2015) 011} [\href{https://arxiv.org/abs/1506.05976}{{\ttfamily
  1506.05976}}].

\bibitem{2017arXiv170201764I}
V.~{Ir{\v s}i{\v c}}, M.~{Viel}, M.~G. {Haehnelt}, J.~S. {Bolton},
  S.~{Cristiani}, G.~D. {Becker} et~al., \emph{{New Constraints on the
  free-streaming of warm dark matter from intermediate and small scale
  Lyman-$\alpha$ forest data}}, {\emph{ArXiv e-prints} (2017) }
  [\href{https://arxiv.org/abs/1702.01764}{{\ttfamily 1702.01764}}].

\bibitem{2017arXiv170203314Y}
C.~{Yeche}, N.~{Palanque-Delabrouille}, J.~{.~Baur} and H.~{du Mas des
  BourBoux}, \emph{{Constraints on neutrino masses from Lyman-alpha forest
  power spectrum with BOSS and XQ-100}}, {\emph{ArXiv e-prints} (2017) }
  [\href{https://arxiv.org/abs/1702.03314}{{\ttfamily 1702.03314}}].

\bibitem{2017arXiv170304683I}
V.~{Ir{\v s}i{\v c}}, M.~{Viel}, M.~G. {Haehnelt}, J.~S. {Bolton} and G.~D.
  {Becker}, \emph{{First constraints on fuzzy dark matter from Lyman-$\alpha$
  forest data and hydrodynamical simulations}}, {\emph{ArXiv e-prints} (2017) }
  [\href{https://arxiv.org/abs/1703.04683}{{\ttfamily 1703.04683}}].

\bibitem{2017arXiv170309126A}
E.~{Armengaud}, N.~{Palanque-Delabrouille}, C.~{Y{\`e}che}, D.~J.~E. {Marsh}
  and J.~{Baur}, \emph{{Constraining the mass of light bosonic dark matter
  using SDSS Lyman-$\alpha$ forest}}, {\emph{ArXiv e-prints} (2017) }
  [\href{https://arxiv.org/abs/1703.09126}{{\ttfamily 1703.09126}}].

\bibitem{2011JCAP...09..001S}
A.~{Slosar}, A.~{Font-Ribera}, M.~M. {Pieri}, J.~{Rich}, J.-M. {Le Goff},
  {\'E}.~{Aubourg} et~al., \emph{{The Lyman-{$\alpha$} forest in three
  dimensions: measurements of large scale flux correlations from BOSS 1st-year
  data}}, \href{https://doi.org/10.1088/1475-7516/2011/09/001}{\emph{\jcap}
  {\bfseries 9} (2011) 001} [\href{https://arxiv.org/abs/1104.5244}{{\ttfamily
  1104.5244}}].

\bibitem{2013A&A...552A..96B}
N.~G. {Busca}, T.~{Delubac}, J.~{Rich}, S.~{Bailey}, A.~{Font-Ribera},
  D.~{Kirkby} et~al., \emph{{Baryon acoustic oscillations in the Ly{$\alpha$}
  forest of BOSS quasars}},
  \href{https://doi.org/10.1051/0004-6361/201220724}{\emph{\aap} {\bfseries
  552} (2013) A96} [\href{https://arxiv.org/abs/1211.2616}{{\ttfamily
  1211.2616}}].

\bibitem{2013JCAP...03..024K}
D.~{Kirkby}, D.~{Margala}, A.~{Slosar}, S.~{Bailey}, N.~G. {Busca},
  T.~{Delubac} et~al., \emph{{Fitting methods for baryon acoustic oscillations
  in the Lyman-{$\alpha$} forest fluctuations in BOSS data release 9}},
  \href{https://doi.org/10.1088/1475-7516/2013/03/024}{\emph{\jcap} {\bfseries
  3} (2013) 024} [\href{https://arxiv.org/abs/1301.3456}{{\ttfamily
  1301.3456}}].

\bibitem{2013JCAP...04..026S}
A.~{Slosar}, V.~{Ir{\v s}i{\v c}}, D.~{Kirkby}, S.~{Bailey}, N.~G. {Busca},
  T.~{Delubac} et~al., \emph{{Measurement of baryon acoustic oscillations in
  the Lyman-{$\alpha$} forest fluctuations in BOSS data release 9}},
  \href{https://doi.org/10.1088/1475-7516/2013/04/026}{\emph{\jcap} {\bfseries
  4} (2013) 026} [\href{https://arxiv.org/abs/1301.3459}{{\ttfamily
  1301.3459}}].

\bibitem{2015A&A...574A..59D}
T.~{Delubac}, J.~E. {Bautista}, N.~G. {Busca}, J.~{Rich}, D.~{Kirkby},
  S.~{Bailey} et~al., \emph{{Baryon acoustic oscillations in the Ly{$\alpha$}
  forest of BOSS DR11 quasars}},
  \href{https://doi.org/10.1051/0004-6361/201423969}{\emph{\aap} {\bfseries
  574} (2015) A59} [\href{https://arxiv.org/abs/1404.1801}{{\ttfamily
  1404.1801}}].

\bibitem{2017arXiv170200176B}
J.~E. {Bautista}, N.~G. {Busca}, J.~{Guy}, J.~{Rich}, M.~{Blomqvist}, H.~{du
  Mas des Bourboux} et~al., \emph{{Measurement of baryon acoustic oscillation
  correlations at z = 2.3 with SDSS DR12 Ly{$\alpha$}-Forests}},
  \href{https://doi.org/10.1051/0004-6361/201730533}{\emph{\aap} {\bfseries
  603} (2017) A12} [\href{https://arxiv.org/abs/1702.00176}{{\ttfamily
  1702.00176}}].

\bibitem{1979Natur.281..358A}
C.~{Alcock} and B.~{Paczynski}, \emph{{An evolution free test for non-zero
  cosmological constant}}, \href{https://doi.org/10.1038/281358a0}{\emph{\nat}
  {\bfseries 281} (1979) 358}.

\bibitem{1999ApJ...511L...5H}
L.~{Hui}, A.~{Stebbins} and S.~{Burles}, \emph{{A Geometrical Test of the
  Cosmological Energy Contents Using the Ly{$\alpha$} Forest}},
  \href{https://doi.org/10.1086/311826}{\emph{\apjl} {\bfseries 511} (1999) L5}
  [\href{https://arxiv.org/abs/astro-ph/9807190}{{\ttfamily
  astro-ph/9807190}}].

\bibitem{1999ApJ...518...24M}
P.~{McDonald} and J.~{Miralda-Escud{\'e}}, \emph{{Measuring the Cosmological
  Geometry from the Ly{$\alpha$} Forest along Parallel Lines of Sight}},
  \href{https://doi.org/10.1086/307264}{\emph{\apj} {\bfseries 518} (1999) 24}
  [\href{https://arxiv.org/abs/astro-ph/9807137}{{\ttfamily
  astro-ph/9807137}}].

\bibitem{2003ApJ...585...34M}
P.~{McDonald}, \emph{{Toward a Measurement of the Cosmological Geometry at z
  \~{} 2: Predicting Ly{$\alpha$} Forest Correlation in Three Dimensions and
  the Potential of Future Data Sets}},
  \href{https://doi.org/10.1086/345945}{\emph{\apj} {\bfseries 585} (2003) 34}
  [\href{https://arxiv.org/abs/astro-ph/0108064}{{\ttfamily
  astro-ph/0108064}}].

\bibitem{1998MNRAS.298L..21H}
M.~G. {Haehnelt} and M.~{Steinmetz}, \emph{{Probing the thermal history of the
  intergalactic medium with Lyalpha absorption lines}},
  \href{https://doi.org/10.1046/j.1365-8711.1998.01879.x}{\emph{\mnras}
  {\bfseries 298} (1998) L21}
  [\href{https://arxiv.org/abs/astro-ph/9706296}{{\ttfamily
  astro-ph/9706296}}].

\bibitem{2018arXiv180804367W}
M.~{Walther}, J.~{O{\~n}orbe}, J.~F. {Hennawi} and Z.~{Luki{\'c}}, \emph{{New
  Constraints on IGM Thermal Evolution from the Ly$\{$$\backslash$alpha$\}$
  Forest Power Spectrum}}, {\emph{ArXiv e-prints} (2018) }
  [\href{https://arxiv.org/abs/1808.04367}{{\ttfamily 1808.04367}}].

\bibitem{2014JCAP...05..023F}
A.~{Font-Ribera}, P.~{McDonald}, N.~{Mostek}, B.~A. {Reid}, H.-J. {Seo} and
  A.~{Slosar}, \emph{{DESI and other Dark Energy experiments in the era of
  neutrino mass measurements}},
  \href{https://doi.org/10.1088/1475-7516/2014/05/023}{\emph{\jcap} {\bfseries
  5} (2014) 023} [\href{https://arxiv.org/abs/1308.4164}{{\ttfamily
  1308.4164}}].

\bibitem{2016AJ....151...44D}
K.~S. {Dawson}, J.-P. {Kneib}, W.~J. {Percival}, S.~{Alam}, F.~D. {Albareti},
  S.~F. {Anderson} et~al., \emph{{The SDSS-IV Extended Baryon Oscillation
  Spectroscopic Survey: Overview and Early Data}},
  \href{https://doi.org/10.3847/0004-6256/151/2/44}{\emph{\aj} {\bfseries 151}
  (2016) 44} [\href{https://arxiv.org/abs/1508.04473}{{\ttfamily 1508.04473}}].

\bibitem{2016arXiv161100036D}
{DESI Collaboration}, A.~{Aghamousa}, J.~{Aguilar}, S.~{Ahlen}, S.~{Alam},
  L.~E. {Allen} et~al., \emph{{The DESI Experiment Part I: Science,Targeting,
  and Survey Design}}, {\emph{ArXiv e-prints} (2016) }
  [\href{https://arxiv.org/abs/1611.00036}{{\ttfamily 1611.00036}}].

\bibitem{2016arXiv161100037D}
{DESI Collaboration}, A.~{Aghamousa}, J.~{Aguilar}, S.~{Ahlen}, S.~{Alam},
  L.~E. {Allen} et~al., \emph{{The DESI Experiment Part II: Instrument
  Design}}, {\emph{ArXiv e-prints} (2016) }
  [\href{https://arxiv.org/abs/1611.00037}{{\ttfamily 1611.00037}}].

\bibitem{2014ApJ...792L..34P}
A.~{Pontzen}, S.~{Bird}, H.~{Peiris} and L.~{Verde}, \emph{{Constraints on
  Ionizing Photon Production from the Large-scale Ly{\ensuremath{\alpha}}
  Forest}}, \href{https://doi.org/10.1088/2041-8205/792/2/L34}{\emph{\apj}
  {\bfseries 792} (2014) L34}
  [\href{https://arxiv.org/abs/1407.6367}{{\ttfamily 1407.6367}}].

\bibitem{Sacks_1989}
J.~Sacks, W.~J. Welch, T.~J. Mitchell and H.~P. Wynn, \emph{Design and analysis
  of computer experiments},
  \href{https://doi.org/10.1214/ss/1177012413}{\emph{Statist. Sci.} {\bfseries
  4} (1989) 409}.

\bibitem{Queipo_2005}
N.~Queipo, R.~Haftka, W.~Shyy, T.~Goel, R.~Vaidyanathan and P.~Tucker,
  \emph{Surrogate-based analysis and optimization}, {\emph{Progress in
  Aerospace Sciences} {\bfseries 41} (2005) 1}.

\bibitem{forrester2008engineering}
A.~Forrester, A.~Sobester and A.~Keane, \emph{Engineering Design via Surrogate
  Modelling: A Practical Guide}. Wiley, 2008.

\bibitem{forrester2009recent}
A.~I. Forrester and A.~J. Keane, \emph{Recent advances in surrogate-based
  optimization}, {\emph{Progress in Aerospace Sciences} {\bfseries 45} (2009)
  50}.

\bibitem{OYEBAMIJI201769}
O.~Oyebamiji, D.~Wilkinson, P.~Jayathilake, T.~Curtis, S.~Rushton, B.~Li
  et~al., \emph{Gaussian process emulation of an individual-based model
  simulation of microbial communities},
  \href{https://doi.org/https://doi.org/10.1016/j.jocs.2017.08.006}{\emph{Journal
  of Computational Science} {\bfseries 22} (2017) 69 }.

\bibitem{Heitmann:2009}
K.~{Heitmann}, D.~{Higdon}, M.~{White}, S.~{Habib}, B.~J. {Williams},
  E.~{Lawrence} et~al., \emph{{The Coyote Universe. II. Cosmological Models and
  Precision Emulation of the Nonlinear Matter Power Spectrum}},
  \href{https://doi.org/10.1088/0004-637X/705/1/156}{\emph{\apj} {\bfseries
  705} (2009) 156} [\href{https://arxiv.org/abs/0902.0429}{{\ttfamily
  0902.0429}}].

\bibitem{Kwan:2013}
J.~Kwan, K.~Heitmann, S.~Habib, N.~Padmanabhan, H.~Finkel, E.~Lawrence et~al.,
  \emph{{Cosmic Emulation: Fast Predictions for the Galaxy Power Spectrum}},
  \href{https://doi.org/10.1088/0004-637X/810/1/35}{\emph{\apj} {\bfseries 810}
  (2015) 35} [\href{https://arxiv.org/abs/1311.6444}{{\ttfamily 1311.6444}}].

\bibitem{2018arXiv180405867Z}
Z.~{Zhai}, J.~L. {Tinker}, M.~R. {Becker}, J.~{DeRose}, Y.-Y. {Mao},
  T.~{McClintock} et~al., \emph{{The Aemulus Project III: Emulation of the
  Galaxy Correlation Function}}, {\emph{ArXiv e-prints} (2018) }
  [\href{https://arxiv.org/abs/1804.05867}{{\ttfamily 1804.05867}}].

\bibitem{Liu:2015}
J.~{Liu}, A.~{Petri}, Z.~{Haiman}, L.~{Hui}, J.~M. {Kratochvil} and M.~{May},
  \emph{{Cosmology constraints from the weak lensing peak counts and the power
  spectrum in CFHTLenS data}},
  \href{https://doi.org/10.1103/PhysRevD.91.063507}{\emph{\prd} {\bfseries 91}
  (2015) 063507} [\href{https://arxiv.org/abs/1412.0757}{{\ttfamily
  1412.0757}}].

\bibitem{Petri:2015}
A.~{Petri}, J.~{Liu}, Z.~{Haiman}, M.~{May}, L.~{Hui} and J.~M. {Kratochvil},
  \emph{{Emulating the CFHTLenS weak lensing data: Cosmological constraints
  from moments and Minkowski functionals}},
  \href{https://doi.org/10.1103/PhysRevD.91.103511}{\emph{\prd} {\bfseries 91}
  (2015) 103511} [\href{https://arxiv.org/abs/1503.06214}{{\ttfamily
  1503.06214}}].

\bibitem{2018arXiv181109141J}
W.~D. {Jennings}, C.~A. {Watkinson}, F.~B. {Abdalla} and J.~D. {McEwen},
  \emph{{Evaluating machine learning techniques for predicting power spectra
  from reionization simulations}}, {\emph{ArXiv e-prints} (2018) }
  [\href{https://arxiv.org/abs/1811.09141}{{\ttfamily 1811.09141}}].

\bibitem{2018arXiv180405866M}
T.~{McClintock}, E.~{Rozo}, M.~R. {Becker}, J.~{DeRose}, Y.-Y. {Mao},
  S.~{McLaughlin} et~al., \emph{{The Aemulus Project II: Emulating the Halo
  Mass Function}}, {\emph{ArXiv e-prints} (2018) }
  [\href{https://arxiv.org/abs/1804.05866}{{\ttfamily 1804.05866}}].

\bibitem{2018PhRvD..98h3540M}
R.~{Murgia}, V.~{Ir{\v{s}}i{\v{c}}} and M.~{Viel}, \emph{{Novel constraints on
  noncold, nonthermal dark matter from Lyman-{\ensuremath{\alpha}} forest
  data}}, \href{https://doi.org/10.1103/PhysRevD.98.083540}{\emph{\prd}
  {\bfseries 98} (2018) 083540}
  [\href{https://arxiv.org/abs/1806.08371}{{\ttfamily 1806.08371}}].

\bibitem{webster2007geostatistics}
R.~Webster and M.~A. Oliver, \emph{Geostatistics for environmental scientists}.
  John Wiley \& Sons, 2007.

\bibitem{2006MNRAS.365..231V}
M.~{Viel} and M.~G. {Haehnelt}, \emph{{Cosmological and astrophysical
  parameters from the Sloan Digital Sky Survey flux power spectrum and
  hydrodynamical simulations of the Lyman {$\alpha$} forest}},
  \href{https://doi.org/10.1111/j.1365-2966.2005.09703.x}{\emph{\mnras}
  {\bfseries 365} (2006) 231}
  [\href{https://arxiv.org/abs/astro-ph/0508177}{{\ttfamily
  astro-ph/0508177}}].

\bibitem{2011MNRAS.413.1717B}
S.~{Bird}, H.~V. {Peiris}, M.~{Viel} and L.~{Verde}, \emph{{Minimally
  parametric power spectrum reconstruction from the Lyman {$\alpha$} forest}},
  \href{https://doi.org/10.1111/j.1365-2966.2011.18245.x}{\emph{\mnras}
  {\bfseries 413} (2011) 1717}
  [\href{https://arxiv.org/abs/1010.1519}{{\ttfamily 1010.1519}}].

\bibitem{2013A&A...559A..85P}
N.~{Palanque-Delabrouille}, C.~{Y{\`e}che}, A.~{Borde}, J.-M. {Le Goff},
  G.~{Rossi}, M.~{Viel} et~al., \emph{{The one-dimensional Ly{$\alpha$} forest
  power spectrum from BOSS}},
  \href{https://doi.org/10.1051/0004-6361/201322130}{\emph{\aap} {\bfseries
  559} (2013) A85} [\href{https://arxiv.org/abs/1306.5896}{{\ttfamily
  1306.5896}}].

\bibitem{2015JCAP...02..045P}
N.~{Palanque-Delabrouille}, C.~{Y{\`e}che}, J.~{Lesgourgues}, G.~{Rossi},
  A.~{Borde}, M.~{Viel} et~al., \emph{{Constraint on neutrino masses from
  SDSS-III/BOSS Ly{$\alpha$} forest and other cosmological probes}},
  \href{https://doi.org/10.1088/1475-7516/2015/02/045}{\emph{\jcap} {\bfseries
  2} (2015) 045} [\href{https://arxiv.org/abs/1410.7244}{{\ttfamily
  1410.7244}}].

\bibitem{emulator_forest_GP_2018}
S.~{Bird}, K.~K. {Rogers}, H.~V. {Peiris}, L.~{Verde}, A.~{Font-Ribera} and
  A.~{Pontzen}, \emph{{An Emulator for the Lyman-alpha Forest}}, {\emph{arXiv
  e-prints} (2018) arXiv:1812.04654}
  [\href{https://arxiv.org/abs/1812.04654}{{\ttfamily 1812.04654}}].

\bibitem{gpml}
C.~E. Rasmussen and C.~K.~I. Williams, \emph{{Gaussian Processes for Machine
  Learning}}. MIT Press, 2006.

\bibitem{kushner1964new}
H.~J. Kushner, \emph{A new method of locating the maximum point of an arbitrary
  multipeak curve in the presence of noise}, {\emph{Journal of Basic
  Engineering} {\bfseries 86} (1964) 97}.

\bibitem{mockus1978toward}
J.~Mockus, V.~Tiesis and A.~Zilinskas, \emph{Toward global optimization, volume
  2, chapter bayesian methods for seeking the extremum}, .

\bibitem{Mockus1994}
J.~Mockus, \emph{Application of bayesian approach to numerical methods of
  global and stochastic optimization},
  \href{https://doi.org/10.1007/BF01099263}{\emph{Journal of Global
  Optimization} {\bfseries 4} (1994) 347}.

\bibitem{10.2307/2673557}
M.~C. Kennedy and A.~O'Hagan, \emph{Predicting the output from a complex
  computer code when fast approximations are available}, {\emph{Biometrika}
  {\bfseries 87} (2000) 1}.

\bibitem{kennedy2001bayesian}
M.~C. Kennedy and A.~O'Hagan, \emph{Bayesian calibration of computer models},
  {\emph{Journal of the Royal Statistical Society: Series B (Statistical
  Methodology)} {\bfseries 63} (2001) 425}.

\bibitem{2018PhRvD..98f3511L}
F.~{Leclercq}, \emph{{Bayesian optimization for likelihood-free cosmological
  inference}}, \href{https://doi.org/10.1103/PhysRevD.98.063511}{\emph{\prd}
  {\bfseries 98} (2018) 063511}
  [\href{https://arxiv.org/abs/1805.07152}{{\ttfamily 1805.07152}}].

\bibitem{2014A&A...568A..22B}
M.~{Betoule}, R.~{Kessler}, J.~{Guy}, J.~{Mosher}, D.~{Hardin}, R.~{Biswas}
  et~al., \emph{{Improved cosmological constraints from a joint analysis of the
  SDSS-II and SNLS supernova samples}},
  \href{https://doi.org/10.1051/0004-6361/201423413}{\emph{\aap} {\bfseries
  568} (2014) A22} [\href{https://arxiv.org/abs/1401.4064}{{\ttfamily
  1401.4064}}].

\bibitem{Locatelli1997}
M.~Locatelli, \emph{Bayesian algorithms for one-dimensional global
  optimization}, \href{https://doi.org/10.1023/A:1008294716304}{\emph{Journal
  of Global Optimization} {\bfseries 10} (1997) 57}.

\bibitem{Cox97sdo:a}
D.~D. Cox and S.~John, \emph{Sdo: A statistical method for global
  optimization},  in \emph{in Multidisciplinary Design Optimization:
  State-of-the-Art}, pp.~315--329, 1997.

\bibitem{auer2002using}
P.~Auer, \emph{Using confidence bounds for exploitation-exploration
  trade-offs}, {\emph{Journal of Machine Learning Research} {\bfseries 3}
  (2002) 397}.

\bibitem{auer2002finite}
P.~Auer, N.~Cesa-Bianchi and P.~Fischer, \emph{Finite-time analysis of the
  multiarmed bandit problem}, {\emph{Machine learning} {\bfseries 47} (2002)
  235}.

\bibitem{dani2008stochastic}
V.~Dani, T.~P. Hayes and S.~M. Kakade, \emph{Stochastic linear optimization
  under bandit feedback}, .

\bibitem{2017arXiv170400520J}
M.~{J{\"a}rvenp{\"a}{\"a}}, M.~U. {Gutmann}, A.~{Pleska}, A.~{Vehtari} and
  P.~{Marttinen}, \emph{{Efficient acquisition rules for model-based
  approximate Bayesian computation}}, {\emph{ArXiv e-prints} (2017)
  arXiv:1704.00520} [\href{https://arxiv.org/abs/1704.00520}{{\ttfamily
  1704.00520}}].

\bibitem{yu_feng_2018_1451799}
Y.~Feng, S.~Bird, L.~Anderson, A.~Font-Ribera and C.~Pedersen,
  \emph{Mp-gadget/mp-gadget: A tag for getting a doi},  Oct., 2018.
\newblock 10.5281/zenodo.1451799.

\bibitem{2001NewA....6...79S}
V.~{Springel}, N.~{Yoshida} and S.~D.~M. {White}, \emph{{GADGET: a code for
  collisionless and gasdynamical cosmological simulations}},
  \href{https://doi.org/10.1016/S1384-1076(01)00042-2}{\emph{\na} {\bfseries 6}
  (2001) 79} [\href{https://arxiv.org/abs/astro-ph/0003162}{{\ttfamily
  astro-ph/0003162}}].

\bibitem{2005MNRAS.364.1105S}
V.~{Springel}, \emph{{The cosmological simulation code GADGET-2}},
  \href{https://doi.org/10.1111/j.1365-2966.2005.09655.x}{\emph{\mnras}
  {\bfseries 364} (2005) 1105}
  [\href{https://arxiv.org/abs/astro-ph/0505010}{{\ttfamily
  astro-ph/0505010}}].

\bibitem{2011AJ....142...72E}
D.~J. {Eisenstein}, D.~H. {Weinberg}, E.~{Agol}, H.~{Aihara}, C.~{Allende
  Prieto}, S.~F. {Anderson} et~al., \emph{{SDSS-III: Massive Spectroscopic
  Surveys of the Distant Universe, the Milky Way, and Extra-Solar Planetary
  Systems}}, \href{https://doi.org/10.1088/0004-6256/142/3/72}{\emph{\aj}
  {\bfseries 142} (2011) 72} [\href{https://arxiv.org/abs/1101.1529}{{\ttfamily
  1101.1529}}].

\bibitem{2013AJ....145...10D}
K.~S. {Dawson}, D.~J. {Schlegel}, C.~P. {Ahn}, S.~F. {Anderson},
  {\'E}.~{Aubourg}, S.~{Bailey} et~al., \emph{{The Baryon Oscillation
  Spectroscopic Survey of SDSS-III}},
  \href{https://doi.org/10.1088/0004-6256/145/1/10}{\emph{\aj} {\bfseries 145}
  (2013) 10} [\href{https://arxiv.org/abs/1208.0022}{{\ttfamily 1208.0022}}].

\bibitem{2017ascl.soft10012B}
S.~{Bird}, ``{FSFE: Fake Spectra Flux Extractor}.'' Astrophysics Source Code
  Library, Oct., 2017.

\bibitem{2008MNRAS.386.1131B}
J.~S. {Bolton}, M.~{Viel}, T.~S. {Kim}, M.~G. {Haehnelt} and R.~F. {Carswell},
  \emph{{Possible evidence for an inverted temperature-density relation in the
  intergalactic medium from the flux distribution of the
  Ly{\ensuremath{\alpha}} forest}},
  \href{https://doi.org/10.1111/j.1365-2966.2008.13114.x}{\emph{\mnras}
  {\bfseries 386} (2008) 1131}
  [\href{https://arxiv.org/abs/0711.2064}{{\ttfamily 0711.2064}}].

\bibitem{Kim:2007}
T.~{Kim}, J.~S. {Bolton}, M.~{Viel}, M.~G. {Haehnelt} and R.~F. {Carswell},
  \emph{{An improved measurement of the flux distribution of the Ly{$\alpha$}
  forest in QSO absorption spectra: the effect of continuum fitting, metal
  contamination and noise properties}},
  \href{https://doi.org/10.1111/j.1365-2966.2007.12406.x}{\emph{\mnras}
  {\bfseries 382} (2007) 1657}
  [\href{https://arxiv.org/abs/0711.1862}{{\ttfamily 0711.1862}}].

\bibitem{2002PhRvD..66j3511L}
A.~{Lewis} and S.~{Bridle}, \emph{{Cosmological parameters from CMB and other
  data: A Monte Carlo approach}},
  \href{https://doi.org/10.1103/PhysRevD.66.103511}{\emph{\prd} {\bfseries 66}
  (2002) 103511} [\href{https://arxiv.org/abs/astro-ph/0205436}{{\ttfamily
  astro-ph/0205436}}].

\bibitem{2013PASP..125..306F}
D.~{Foreman-Mackey}, D.~W. {Hogg}, D.~{Lang} and J.~{Goodman}, \emph{{emcee:
  The MCMC Hammer}}, \href{https://doi.org/10.1086/670067}{\emph{\pasp}
  {\bfseries 125} (2013) 306}
  [\href{https://arxiv.org/abs/1202.3665}{{\ttfamily 1202.3665}}].

\bibitem{srinivas2009gaussian}
N.~Srinivas, A.~Krause, S.~M. Kakade and M.~Seeger, \emph{Gaussian process
  optimization in the bandit setting: No regret and experimental design},
  {\emph{arXiv preprint arXiv:0912.3995} (2009) }.

\bibitem{2010arXiv1012.2599B}
E.~{Brochu}, V.~M. {Cora} and N.~{de Freitas}, \emph{{A Tutorial on Bayesian
  Optimization of Expensive Cost Functions, with Application to Active User
  Modeling and Hierarchical Reinforcement Learning}}, {\emph{ArXiv e-prints}
  (2010) } [\href{https://arxiv.org/abs/1012.2599}{{\ttfamily 1012.2599}}].

\bibitem{2015arXiv150103291G}
M.~U. {Gutmann} and J.~{Corander}, \emph{{Bayesian Optimization for
  Likelihood-Free Inference of Simulator-Based Statistical Models}},
  {\emph{ArXiv e-prints} (2015) }
  [\href{https://arxiv.org/abs/1501.03291}{{\ttfamily 1501.03291}}].

\bibitem{sobol1967distribution}
I.~M. Sobol', \emph{On the distribution of points in a cube and the approximate
  evaluation of integrals}, {\emph{Zhurnal Vychislitel'noi Matematiki i
  Matematicheskoi Fiziki} {\bfseries 7} (1967) 784}.

\end{thebibliography}\endgroup







\end{document}